# Surface charge density of current-carrying conductors: An exact analytical solution for infinitely thin wires of arbitrary shape

R. Merlin

*The Harrison M. Randall Laboratory of Physics, University of Michigan, Ann Arbor, Michigan 48109-1040, USA*

An exact asymptotic expression is derived for the surface charge density associated with a steady current in closed loops or wire segments of infinitesimal cross section connected to a battery. Except for vanishingly thin boundary layers at the ends of the wire, and irrespective of the conductor's shape, the charge density varies linearly with arc length, as does the electrostatic potential along the curve. This proportionality generalizes earlier results and provides a clear physical picture of charge accumulation in electrical circuits.

The transition from electrostatics to electrical circuits in introductory electricity and magnetism courses can feel a bit disconcerting. After spending several weeks calculating the surface charges that shield external fields — so that the field inside a conductor is zero — students are suddenly asked to set that picture aside and consider situations in which the internal field is nonvanishing. Only occasionally do they hear, and often with some surprise, that the field driving the current actually comes from charges on the surface of the conductor, not from the battery itself. With a few exceptions [1,2,3,4,5], both undergraduate and graduate textbooks tend to sidestep these issues.

The question of surface charges and, consequently, of the electric field outside a conductor carrying a steady current, has a notable, though often underemphasized, place in the historical development of electromagnetism. The problem has been taken up by many authors and has resurfaced repeatedly over the years [6,7,8]. Analytical solutions are known for a few simple geometries, including a long straight wire [1,9,10,11], a cylindrical hollow conductor with a circulating sheet current [12], coaxial cables [13], and a toroidal conductor carrying an azimuthal current [14,15]. In this paper, we provide an exact asymptotic solution for the surface charge distribution along a wire of arbitrary shape and infinitesimal cross section carrying a steady current. The approach adopted here is related to that used in a prior electrostatic analysis of the field produced by a one-dimensional charge distribution on a curve [16].

Consider a cylindrical, ohmic conductor whose axis follows a (closed or open) curve $\mathbf{r}(s)$ of length $L$, where $\mathbf{r}=(x,y,z)$ and $s\in(0,L)$ denotes the arc length, with a circular cross section of radius $a$ in the plane normal to the curve at each point; see Fig. 1. We are interested in the limit $a\to 0$. The curve is assumed to be smooth (continuously differentiable), with bounded curvature $\kappa(s) = |d^2\mathbf{r}/ds^2| < \infty$ everywhere and, importantly, to admit a non-self-intersecting tubular neighborhood of radius $R >> a$, that is, it can be thickened into a tube of radius $R$ without different parts of the tube overlapping. This ensures that distinct segments of the wire remain well separated on the scale of its thickness. Topologically, the curve is equivalent to a ring if closed, and to a segment if open.

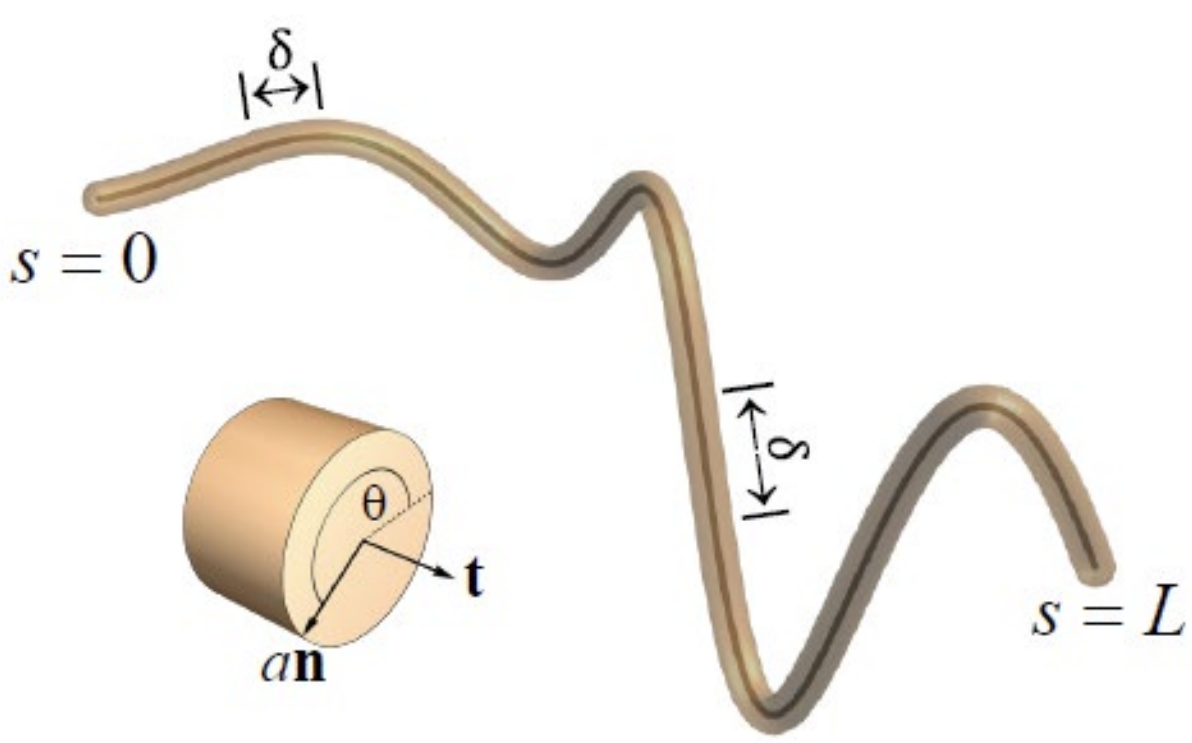

**Fig. 1** Idealized representation of a thin wire modeled as a cylindrical tube (the actual radius is much smaller than that depicted in the figure). The black line denotes the parametric curve $\mathbf{r}(s)$ defining the wire axis. The tube is generated by transporting a circular cross section of radius $a$ along this curve. The parameter $\delta >> a$ introduced in the text is schematically shown for two values of $s$. Owing to the assumption that the wire can be thickened into a non-self-intersecting tube of radius $R >> a$, points outside the near region defined by $\delta$ are separated from the reference point by a Euclidean distance much larger than $a$. Unit vectors $\mathbf{t}$ and $\mathbf{n}$, and the azimuthal angle $\theta$ are shown in the inset.

The wire is connected to a battery via leads. In the case of a closed loop, the battery introduces a potential discontinuity at $\mathbf{r}(L) = \mathbf{r}(0)$. A surface charge density $\Sigma(s,\theta)$ appears on the wire such that the tangential electric field inside the conductor has constant magnitude $E_0$ ($\theta$ is the azimuthal angle in the plane normal to the curve) [17]. At an arbitrary point on the axis, the charge distribution is then determined as the solution to the integral equation (Gaussian units)

$$V(s) = -E_0 s = \int_0^L \int_0^{2\pi} \frac{\Sigma(s',\theta') a ds' d\theta'}{\left|\mathbf{r}(s) - \mathbf{r}(s') - a\mathbf{n}(s',\theta')\right|} \quad , \tag{1}$$

where $\mathbf{n}(s,\theta)$ is a unit vector in the cross-sectional plane, as shown in Fig. 1. For $a = 0$, the potential exhibits a non-integrable singularity at $s' = s$; the finite radius $a > 0$ regularizes this expression.

To proceed, we take $a$ to be much smaller than the lower bound of the inverse curvature, and split the integral over $s'$ into the near region $|s'-s|<\delta(s)$ and the non-singular far regions $|s'-s|\geq\delta$, where $a \ll \delta \ll 1/\kappa(s)$. A parameter satisfying these inequalities can always be found because the microscopic scale, $a$, is well separated from the curvature scale, $1/\kappa_0$.

In the near region, we expand around $s'=s$ over a segment that is much shorter than the local radius of curvature. Using $\mathbf{t}=d\mathbf{r}/ds$, we have $\mathbf{r}(s)-\mathbf{r}(s')-a\mathbf{n}(s',\theta)=-\Delta s\mathbf{t}-a\mathbf{n}(s,\theta)-a\Delta s\partial_s\mathbf{n}+O(\Delta s^2)$, where $\Delta s=s'-s$. Since $|\partial_s\mathbf{t}|\equiv\kappa$, $|\partial_s^2\mathbf{n}|\sim|\kappa\,\partial_s\mathbf{n}|$ and $\kappa|\Delta s|<\kappa\delta \ll 1$, terms quadratic in $\Delta s$ can be neglected. Moreover, we can also neglect the term on $\partial_s\mathbf{n}$ because $|\partial_s\mathbf{n}|/\kappa\sim O(1)$ and, thus, $a\Delta s|\partial_s\mathbf{n}|\leq a\delta/\kappa \ll a$. Therefore, $|\mathbf{r}(s)-\mathbf{r}(s')-a\mathbf{n}(s',\theta)|^2\approx\Delta s^2+a^2$.

In the far region, $|s'-s|\geq\delta \gg a$. Given that the curve admits a non-self-intersecting neighborhood of radius $R \gg a$, at intermediate and large separation, the Euclidean distance is at least $R \gg a$, whereas for $|s'-s| \ll R$, $|\mathbf{r}(s)-\mathbf{r}(s')|$ is of the same order as $|s'-s|$. Accordingly, $|\mathbf{r}(s)-\mathbf{r}(s')-a\mathbf{n}(s',\theta)|\approx|\mathbf{r}(s)-\mathbf{r}(s')|$, which can in turn be approximated by $\left[|\mathbf{r}(s)-\mathbf{r}(s')|^2+a^2\right]^{1/2}$, allowing the far-region expression to be cast into the same form as the near-region contribution. Thus, to a high degree of accuracy for all $s'$, the potential can be written compactly as

$$V(s)\approx\int_0^L\frac{\lambda(s')ds'}{\sqrt{|\mathbf{r}(s)-\mathbf{r}(s')|^2+a^2}}\quad, \tag{2}$$

where $\lambda(s')=a\int\Sigma(s',\theta')d\theta'$ is the linear charge density.

To isolate the effect of the singularity, we add and subtract from $V(s)$ the term

$$\Xi(s)=\int_0^L \frac{\lambda(s')ds'}{\sqrt{(s-s')^2+a^2}} \quad , \tag{3}$$

which is the expression for the potential of a straight segment. The crucial step is to show that the singular behavior of $V$ is exactly cancelled by $\Xi$, so that $|V(s)-\Xi(s)|$ remains bounded when $a\to 0$. To this end, it suffices to evaluate this expression in the near region, since $V$ and $\Xi$ have no singularities in the far region. Expanding the integrand about $s$, we have

$$\lim_{a\to 0}\left\{\int_{|s-s'|<\delta}\left[\frac{1}{\sqrt{|\mathbf{r}(s)-\mathbf{r}(s')|^2+a^2}}-\frac{1}{|\sqrt{(s-s')^2+a^2}}\right]\lambda(s')ds'\right\}\approx -\frac{\delta^2}{8}|\kappa(s)|^2\,\lambda(s) \quad . \tag{4}$$

With the singularity removed, $|V(s)-\Xi(s)|$ is then finite for all $s$. Hence, for $a\to 0$, $V(s)$ and $\Xi(s)$ have the same leading-order divergent behavior, thereby reducing the general case of a curved wire onto that of a linear segment.

The factor multiplying $\lambda(s')$ in Eq. (3) is strongly peaked at $s'=s$. This suggests that the integral in Eq. (3) is dominated by the local contribution as $a\to 0$, which should result in $\Xi(s)\propto\lambda(s)$. For $V=-E_0 s$, the asymptotic solution to the integral equation should thus be $\lambda(s)\propto s$. This heuristic argument can be verified by first performing the straightforward integration, yielding

$$\Xi(s)\propto\int_0^L \frac{s'ds'}{\sqrt{(s-s')^2+a^2}}=\sqrt{(s-L)^2+a^2}-\sqrt{s^2+a^2}+s\,\Phi(s) \quad , \tag{5}$$

where

$$\Phi(s)=-\log\left[\frac{s-L+\sqrt{(s-L)^2+a^2}}{s+\sqrt{s^2+a^2}}\right]\approx\log\left[4s(L-s)/a^2\right] \tag{6}$$

for $s >> a$ and $L - s >> a$. Multiplying and dividing the argument of the logarithm by $a^2 / L^2$, $\Phi(s) \approx \log\left[4\varepsilon(1-\varepsilon)\right] + \log(L^2 / a^2)$, where $\varepsilon = s / L$. Thus, if $\left|\log\left[4\varepsilon(1-\varepsilon)\right]\right| << \left|\log(L^2 / a^2)\right|$, $\Phi(s)$ is up to negligible corrections a constant equal to $\log(L^2 / a^2)$.

Now, in the range

$$(a / L)^{\eta} < \varepsilon < 1 - (a / L)^{\eta} \quad , \tag{7}$$

where $0 < \eta < 1$ is required under the constraints $0 < \varepsilon < 1$, $s >> a$ and $L - s >> a$,

$$-\log\left[\varepsilon(1-\varepsilon)\right] < \eta \log(L^2 / a^2) < \log(L^2 / a^2) \quad . \tag{8}$$

For $0 < \eta < 1$, this expression is valid when $s >> a$ and $L - s >> a$, which are precisely the conditions under which the approximation $\Phi(s) \approx \log\left[4s(L-s) / a^2\right]$ applies. It is then apparent from Eq. (8) that the constancy of $\Phi$ is satisfied whenever $\eta << 1$. Since, for fixed $\eta$, the range defined in Eq. (7) slowly covers the whole wire as $a / L$ approaches zero, $\Phi$ becomes asymptotically equal to $\log(L^2 / a^2)$ along the whole curve. The plots of the exact expression of $\Phi(s)$, shown in Fig. 2 for different values of $a / L$, clearly demonstrate that $\Phi(s)$ tends toward a constant over an increasingly wider range as $a / L \to 0$.

From Eq. (5), $d\Xi / ds$ and thus, the electric field, is proportional to

$$\frac{(s-L)}{\sqrt{(s-L)^2 + a^2}} - \frac{s}{\sqrt{s^2 + a^2}} - \Phi - s d\Phi / ds \quad , \tag{9}$$

which, for $s >> a$, $L - s >> a$ and $a \to 0$ becomes approximately equal to

$$-\log[4s(L-s) / a^2] - \frac{L - 2s}{L - s} \quad . \tag{10}$$

The boundary effects set in when $\Phi = \log(L^2 / a^2) \sim (L - 2s) / (L - s)$. Thus, except for a region of width $\ell \sim L / \log(L / a) >> a$ near $s = L$, the asymptotic solution to Eq. (1) for $a / L \to 0$ is

$$\lambda(s) = s\frac{E_0}{\log(a^2/L^2)} \quad . \tag{11}$$

This expression, which constitutes the key result of this paper, reveals the expected singular behavior in the vanishing-radius limit. To ensure that the linear charge density λ remains finite, one must take $E_0 \propto \log(a^2/L^2)$. Thus, as the cross-sectional area $A = \pi a^2$ approaches zero, the current $I = \sigma E_0 A$ vanishes (σ is the conductivity). Consequently, one cannot simultaneously maintain a finite current and a finite charge density in an infinitely thin wire. In practical terms, however, the results obtained here are valid for a small but finite radius. Since a uniform charge density does not produce an electric field [16,18,19], had we chosen $\lambda(0) \neq 0$, the modified potential would be linear everywhere except within boundary layers of length $\ell \sim L/\log(L/a)$ near both ends of the wire. It is in these regions that charge must accumulate in order to sustain a constant electric field and, consequently, a steady current. Such charge accumulation near the battery leads is a well-known feature in straight-wires [13] and toroidal geometries [14].

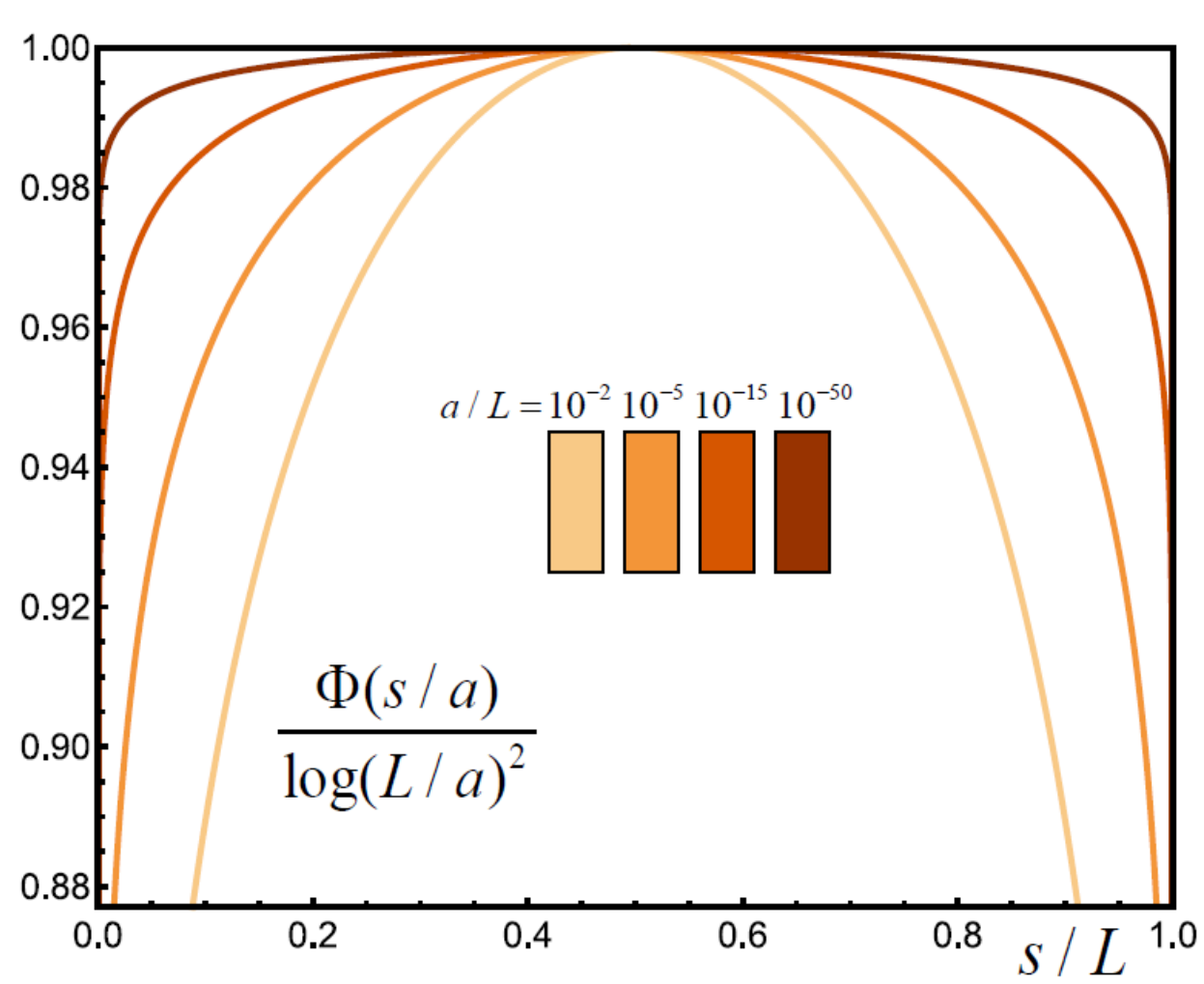

**Figure 2** Normalized Φ(*s*/*a*) as a function of *s*/*L* for different values of *a*/*L*. The plots illustrate how Φ approaches an almost constant profile over the entire domain as $a/L \to 0$, together with the progressive reduction of boundary-induced deviations.

It is worth noting that, unlike wires of vanishingly small cross-sectional area, the behavior of cylinders carrying an azimuthal current is not universal. To understand this difference, consider the two dimensional counterpart to Eq. (1)

$$-E_0 s = 2\int_0^L \log|\boldsymbol{\rho}(s)-\boldsymbol{\rho}(s')|^{-1}\,\Sigma(s')ds' \quad , \tag{12}$$

where $\boldsymbol{\rho}=(x,y)$. Contrasting the case of a one-dimensional conductor, where nearby charges dominate because of the non-integrable nature of $|\mathbf{r}(s)-\mathbf{r}(s')|^{-1}$, the logarithmic kernel in this expression, although singular, is fully integrable. Therefore, the charge density that gives the desired uniform field must depend on the shape of the cross section (for a circular cylinder, $\Sigma \propto \tan\varphi$, with $\varphi$ the azimuthal angle [12]).

In summary, we have obtained an exact asymptotic solution for the surface charge distribution in a current-carrying wire of infinitesimal cross section, applicable to simple, smooth curves of arbitrary shape. It generalizes earlier findings for long straight wires [10] and thin toroids [14], both of which exhibit the same linear variation along the current path. We note that arguments analogous to those presented here immediately yield the previously known result that $d\lambda/ds=0$ for $V(s)=V_0$ [16,18]. The presence of a common logarithmic singularity in the corresponding integrals suggests that, more generally, $\lambda(s)\propto s^n$ is the asymptotic solution to $V(s)\propto s^n$. Calculations confirm the conjecture and show that the boundary-layer width decreases as $1/n$ with increasing $n$.

---